\documentclass[aps,prd,natbib,floatfix,noshowkeys,epsfig,graphics]{revtex4}%
\usepackage{graphicx}
\usepackage{amsmath}
\usepackage{amsfonts}
\usepackage{amssymb}
\usepackage{xcolor}
\usepackage{ulem}
\usepackage{multirow}

\newcommand{\newc}{\newcommand}
\newc{\beq}{\begin{equation}}
\newc{\eeq}{\end{equation}}

\newc{\beqa}    {\begin{eqnarray}}
\newc{\eeqa}    {\end{eqnarray}}
\newc{\bs}    {\section}
\newc{\no}    {\\ \nonumber}

\def\mnras{{ Mon. Not. Roy. Astron. Soc.  }}

\def\+{\bar}

\def\bx{{\bf{x}}}

\def\bD{{\bar{D}}}

\newc{\st}    {\stackrel}

\begin{document}

\title{ Final parsec problem of black hole mergers and ultralight dark matter}

\author{Hyeonmo Koo$^{ \tt a \dagger}$, Dongsu Bak$^{\, \tt a,b\dagger}$, Inkyu Park$^{\, \tt a,b}$, Sungwook E. Hong$^{ \tt c,d}$ , Jae-Weon Lee$^{ \tt e}$ }
\email{mike1919@uos.ac.kr,dsbak@uos.ac.kr,icpark@uos.ac.kr, swhong@kasi.re.kr, scikid@jwu.ac.kr}

\thanks{$^\dagger$\,These authors contributed equally to this work.}

\affiliation{ \centerline{\sl a) Physics Department,
University of Seoul, Seoul 02504, Korea}
\centerline{\sl  
b)Natural Science Research Institute,
University of Seoul, 
Seoul 02504,  Korea}
 \centerline{\sl c) 
 Korea Astronomy and Space Science Institute, 
 Daejeon 34055,  Korea}
 \centerline{\sl d) 
 Astronomy Campus, University of Science and Technology, 
 Daejeon 34055,  Korea}
 \centerline{\sl e) 
 Department of Electrical and Electronic Engineering, Jungwon University,
 Chungbuk 28024, Korea}
 }

\date{\today}
\begin{abstract}
 When two galaxies merge, they often produce a supermassive black hole  binary (SMBHB) at their center. Numerical simulations with stars and cold dark matter show that SMBHBs typically stall out at a distance of a few parsecs apart and take billions of years to coalesce. This is known as the final parsec problem.
We suggest that ultralight dark matter (ULDM) halos around SMBHBs can generate dark matter waves due to dynamical friction.  These waves can effectively carry away orbital energy from the black holes, rapidly driving them together.
To test this hypothesis, we performed numerical simulations of black hole binaries inside ULDM halos. 
 Due to gravitational cooling and quasi-normal modes, the loss-cone problem can be avoided. The decay time scale gives lower
bounds on masses of the ULDM particles and SMBHBs comparable to observational data.
Our results imply that ULDM waves can  lead to the rapid orbital decay of black hole binaries.

\end{abstract} 

\maketitle

\section{Introduction}

The mystery of supermassive black hole (SMBH) growth is one of the unsolved problems in astronomy. When two galaxies merge, they can form a supermassive black hole binary (SMBHB) at their center. However, numerical simulations show that SMBHBs typically become stuck at a distance of a few parsecs apart, and can take billions of years to merge. At this distance the density of the stars and gas near the SMBHB is too low for dynamical friction to be efficient, while the loss of orbital energy of the SMBHB due to gravitational waves is only efficient for distances less than $\mathcal{O}(10^{-2})~{\rm pc}$ \cite{Begelman1980, finalpc}. Furthermore, the loss cone is depleted as the black holes (BHs) approach each other. This difficulty is known as the final parsec problem. The gravitational wave background recently observed by NANOGrav \cite{NANOGrav:2023hfp} is usually attributed to efficient SMBH mergers. This fact deepens the mystery. Proposed solutions to the final parsec problem often involve bringing in extra matter, such as additional stars or another BH  interacting with the black hole binaries (BHBs)  to help them merge.

Ultralight (fuzzy) dark matter (ULDM) \cite{1983PhLB..122..221B, 1989PhRvA..39.4207M, sin1, myhalo, 0264-9381-17-1-102, fuzzy} is a promising alternative to cold dark matter (CDM), as it has the potential to solve some of the small-scale issues of CDM such as the missing satellite problem, the plane of satellite galaxies problem, and the core-cusp problem \cite{Park:2022lel, deblok-2002, crisis, Matos:2003pe}.
In this model, the ULDM is in a Bose-Einstein condensate  state of ultralight scalar particles with a typical mass $m \gtrsim 10^{-22} {\rm eV}$. ULDM can be described with a macroscopic wave function $\psi$ and the uncertainty principle suppresses too many small-scale structure formations. Beyond the galactic scale, the ULDM behaves like CDM and hence naturally solves the problems of CDM. This model has also been shown to be able to explain a wide range of astrophysical observations, including the rotation curves of galaxies \cite{0264-9381-17-1-102, Mbelek:2004ff, PhysRevD.69.127502}, and the large-scale structures of the universe \cite{2014NatPh..10..496S}. Recently, there has been a growing interest in the interactions between BHs and ULDM halos surrounding them \cite{superradiance}, as these interactions could change the patterns of gravitational waves generated by BHBs.

In this letter, we suggest that, inside ULDM halos (for a review, see \cite{2009JKPS...54.2622L, 2014ASSP...38..107S, 2014MPLA...2930002R, 2011PhRvD..84d3531C, Marsh:2015xka, Hui:2016ltb}), BHBs can generate dark matter (DM) waves due to dynamical friction (DF) and gravitational cooling (GC). The dynamical friction of ULDM \cite{Lancaster:2019mde} refers to the frictional force that arises from the gravitational interaction between a moving celestial object and ULDM wakes generated by the object. On the other hand, gravitational cooling \cite{gravitationalcooling} is a mechanism for relaxation by ejecting ULDM waves carrying out excessive kinetic energy and momentum. These waves can effectively carry away orbital energy from the BHBs, rapidly driving them together and possibly solving the final parsec problem. To test this hypothesis, we perform numerical simulations of BHBs at $\mathcal{O}({\rm pc})$ separation in an ULDM soliton representing  a central part of the galactic halos. Our results provide some evidence that DM waves could lead to the rapid orbital decay of the BHs.

Although collisonless CDM can temporally exert a larger DF force to SMBHBs compared to ULDM \cite{Boudon:2022dxi, Boey:2024dks}, its relaxation time to refill the phase space, given by 
\beq
t_{\rm relax}(r) = \frac{ N_{CDM}}{8 \ln { N_{CDM}}} \left(\frac{r^3}{G M_{\rm tot}(r)}\right)^{\frac{1}{2}
}  ,
\eeq
can easily exceed the age of the universe \cite{2003MNRAS.339..949R}. In this expression, ${ N_{CDM}}$ is the number of CDM particles, and $M_{\rm tot}(r)$ is the total mass within a radius $r$ from the center of a host galaxy, with $r \simeq 1\,\mathrm{pc}$ in our study. As a result, the collisionless CDM model encounters the same loss-cone depletion problem as stars, making it difficult to resolve the final parsec problem. On the other hand, the collective motion of ULDM particles can help replenish the phase space after interacting with SMBHBs. There are several studies on the effects of DF by ULDM on SMBHBs. These studies typically focus either on the entrance states into DM halos (at $r \gtrsim  {\rm kpc}$) \cite{Boey:2024dks} or  the final merging states (at $r\ll 0.01 {\rm pc}$) \cite{Aurrekoetxea:2023jwk}.

The aim of this paper is to numerically show fast orbital decays of SMBHBs at $r\simeq 1\,{\rm pc}$ due to  ULDM waves, whose dynamics is described by the Schr\"odinger-Poisson equations.
The structure of this paper is organized as follows.
In Sec.~\ref{sec:sim}, we explain our numerical simulation for studying the orbital decay of BHBs.
Then, Sec.~\ref{sec:numerical} and \ref{sec:theoretical} show the numerical results from Sec.~\ref{sec:sim} and their theoretical analyses, respectively.
Finally, Sec.~\ref{sec:summary} summarizes our work.

\section{Simulation}\label{sec:sim}

In our simulation, we shall treat BHs as point particles with the Newtonian gravity.
The post-Newtonian correction to the gravitational potential of a BH at a distance of $r$ from the BH is on the order of $r_s/r$, where $r_s = 2GM_{\rm bh}/c^2$ is the Schwarzschild radius of the black hole. 
For $M_{\rm bh}=10^8 {\rm M}_\odot$ and $r=0.01 \,{\rm pc}$, this correction is $\mathcal{O}(10^{-3})$.
Therefore, the Newtonian approximation of the gravitational field is  good enough for our calculation.

The ULDM system, interacting gravitationally with $N$-particles (BHs), is governed by the Schr\"odinger-Poisson equations
\begin{align}
i\hbar \frac{\partial}{\partial t} \psi(\bx,t ) & = -\frac{\hbar^2}{2m}\nabla^2 \psi(\bx,t )  +m \left[ V_{\rm U}(\bx,t ) +  V_{\rm N}(\bx,t )  \right] \psi(\bx,t )
\label{Schrodinger-Poisson1 eq.}
\\
 \nabla^2 V_{\rm U}(\bx,t )&=4\pi G \, |\psi|^2(\bx,t )
\label{Schrodinger-Poisson eq.}
\end{align}
with the mass $m$ of ultralight DM particles and the Newton constant $G$.
The wave function $\psi(\bx, t)$ for ULDM is a complex function of spacetime normalized as  { $\int {\rm d} \bx\,|\psi|^2 =M$}
with $M$ denoting the total ULDM mass such that the corresponding ULDM mass density is given by $\rho=|\psi|^2$. 
$V_{\rm U}$ is the potential from the ULDM mass distribution and $V_{\rm N}$ for the $N$-particles where $k$-th ($k=1,2,\cdots, N$) particle is located at ${\bf x}_k(t)$ with mass $M_k$.
(In this work, $N=2$.)
Then, the $N$-body potential is given by $V_{\rm N}(\bx,t)=\sum^N_{k=1} V_k(\bx,t)$ with $V_k(\bx,t)= -GM_k /|\bx-\bx_k(t)|$. With these potentials, the remaining $N$-body dynamics is described by
\begin{equation}
\ddot{\bx}_{k} (t)=-\nabla V_{\rm U}(\bx_{k}(t),t)-\sum_{l\ne k} {\nabla V_{l}(\bx_{k}(t),t)} \, .
\label{NBody eq.}
\end{equation}
The total combined system enjoys two exact scaling relations. One is related to the scaling of the total ULDM mass $M$ and the particle mass $M_k$: 
$[\, M,\; M_k,\; m,\; t,\; \bx,\; V_{\rm U},\; V_{k},\; \psi\,] \rightarrow [\, \kappa M,\; \kappa M_k,\; m,\; \kappa^{-2} t,\; \kappa^{-1} \bx,\; \kappa^2 V_{\rm U},\; \kappa^2 V_{k},\;\kappa^2 \psi\,]$. The other is the scaling of the 
ULDM particle mass $m$ with the transformation
$[\, M,\; M_k, \; m,\; t,\; \bx,\; V_{\rm U},\; V_{k},\; \psi\,]\rightarrow [\, M,\; M_k, \; \lambda m,\; \lambda^{-3} t,\;\lambda^{-2} \bx,\; \lambda^2 V_{\rm U},\; \lambda^2 V_{k},\; \lambda^3 \psi]$. Under these two transformations, the forms of the full equations~(\ref{Schrodinger-Poisson1 eq.}), (\ref{Schrodinger-Poisson eq.}) and (\ref{NBody eq.}) remain invariant. These two scaling relations will be useful in the theoretical analysis in  Sec.~\ref{sec:theoretical}.

In our simulation, we use the  Python pseudo-spectral solver, PyUltraLight2 \cite{Wang:2022grqc}, which is publicly available. This package, as in its prior version \cite{Edwards:2018ccc}, solves the Schr\"odinger-Poisson equations but also incorporates the above-mentioned $N$-body dynamics in a full-fledged form. In this package, the system will be placed in a box of size $L$ with a periodic boundary condition $x_i \sim x_i + L$. Below we take the box size to be $L=40 {\rm pc}$, which is large enough to neglect the effect of the periodic boundary condition. The number of grid size in each direction is taken to be $500$, and then the resulting spatial resolution becomes $\Delta x= 0.08 {\rm pc}$.

The default system value of the time resolution is set to be  $\Delta t_d=1.039 \,m^{-1}_{21} {\rm yr}$ with the ULDM particle mass ratio 
$m_{21}=m/(10^{-21} {\rm eV})$, whose validity is rather well tested 
\cite{Wang:2022grqc}. One may introduce the resolution factor (RF) defined by ${\rm RF}=\Delta t/\Delta t_d$, where $\Delta t$ denotes an actual choice of resolution time used for the simulation. Below we  use the two choices: $\Delta t_1$ and $\Delta t_2$ respectively with
${\rm RF1}\approx 0.9821$ and ${\rm RF2}\approx 1.9249$. As will be discussed further below, 
RF1 is only used to see the validity of our simulation in Figure~\ref{energy}. To obtain the time series data of separation and velocity of BHBs and others, we use RF2, which is mainly due to our limitation of computing resources. We have, however, checked that the relative difference of the separations always remains within 1\% during the entire simulation time for one value of the BH mass. 
For the further discussions on the similar choices of time resolution, see \cite{Edwards:2018ccc} and also \cite{Bak:2020det,Park:2022lel}.

In our simulation below, we set the ULDM particle mass
as $m=10^{-21} {\rm eV}$ ($m_{21}=1$) and place a central ULDM halo of mass $M_s=10^9 {\rm M}_\odot$ (hence $M=M_s$) at the center of the box to represent a galactic DM core.
This mass value is somewhat larger than the typical value
($m=10^{-22} {\rm eV}$), but there are recent observations favoring the heavier mass \cite{Zimmermann:2024xvd,Armengaud:2017nkf}.
To clearly see the orbital decay, we chose a rather massive and compact DM soliton than usual one, mainly due to limited simulation resolution and computing resources. 
(This compact region can also be interpreted as the innermost part of the galactic dark matter cloud that surrounds BHBs.)
 In Sec. IV, we theoretically show that a fast orbital decay can be achieved with a more realistic DM density.
The central region will be modeled by the ground state (soliton) configuration  of the Schr\"odinger-Poisson system in the absence of BHs or any other particles. 
The central DM density of the stable soliton core is given by \cite{Hui:2016ltb}
\beq
\rho_0 = 7.05\times 10^6 \left(\frac{m}{10^{-21} \mathrm{eV}}\right)^6\left(\frac{M_s}{10^9 \rm M_{\odot}}\right)^4 \rm M_{\odot}\, \mathrm{pc}^{-3} .
\label{rho0}
\eeq
For $m=10^{-21}\rm eV$ and $M_s=10^9 \mathrm{M}_\odot$ we adopted, this equation gives a density larger than the typical central density of the local dwarf galaxies.
However, almost all SMBHBs are found in massive galaxies, where the central dark matter density can be much higher than that of dwarf galaxies and often has a steeper profile.
For example, the recent S2 orbital data from our galaxy provides an upper bound for the enclosed DM mass $M(r)$ within the S2 orbit, which is about 0.1\% of the SMBH mass ($M_{\rm bh}\simeq 4\times 10^6 \mathrm{M}_\odot$) \cite{Heissel:2021pcw}.
It can be translated into a bound $\rho_0 \lesssim 10^{13} \rm M_{\odot} {\mathrm{pc}}^{-3}$,  which is unacceptably high for a core profile if we saturate the bound. 
Therefore, it is reasonable to expect that the actual DM density near  central SMBHBs in a massive galaxy has a spike-like profile $\rho(r)\propto r^{-\gamma_{\rm sp}}$ with $1<\gamma_{\rm sp}<2$ 
rather than a wide and flat core profile \cite{Gondolo:1999ef,Avilez:2017jql}. Thus, we assume that our initial soliton DM distribution mimics a dense central region in a  core rather than the core itself. 

By analysing the kinematics of a clockwise-rotating disk of stars and the S2 orbit in our galaxy the constraint $M(r=0.3 \mathrm{pc}) \lesssim 10^6 \rm M_{\odot}$ is obtained for ULDM spikes \cite{Bar:2019pnz}. 
A similar bound $M(r=1 \mathrm{pc}) \lesssim 10^5 \rm M_{\odot}$ is obtained for 
collisionless DM models \cite{Lacroix:2018zmg}.
Therefore, for our theoretical analysis we will take $\rho(r = 1 \mathrm{pc}) \lesssim 10^5 \rm M_{\odot}\, pc^{-3}$ as a conservative observational upper bound for the DM density at $1\,\mathrm{pc}$ in this paper. (See Figure 5.)

We  place two BHs of the same mass $M_{\rm bh}$ at $(\pm 0.45 \rm pc,\,0,\,0)$ respectively. To demonstrate  the orbital decay of BHB, we generate $91$  samples with uniformly varying $M_{\rm bh}$ from $0.6\times 10^8 {\rm M}_\odot$ to $1.5\times 10^8 {\rm M}_\odot$. In each sample, the initial velocity of each BH is chosen such that the corresponding initial orbit becomes nearly circular and the total simulation time set to be $T_{\rm sim}=100 \, m^{-1}_{21}\,\rm kyr$.

\section{Numerical results}\label{sec:numerical}

\begin{figure*}[tb]
\centering
\includegraphics[width=\textwidth,trim={4cm 1cm 5cm 1cm}]{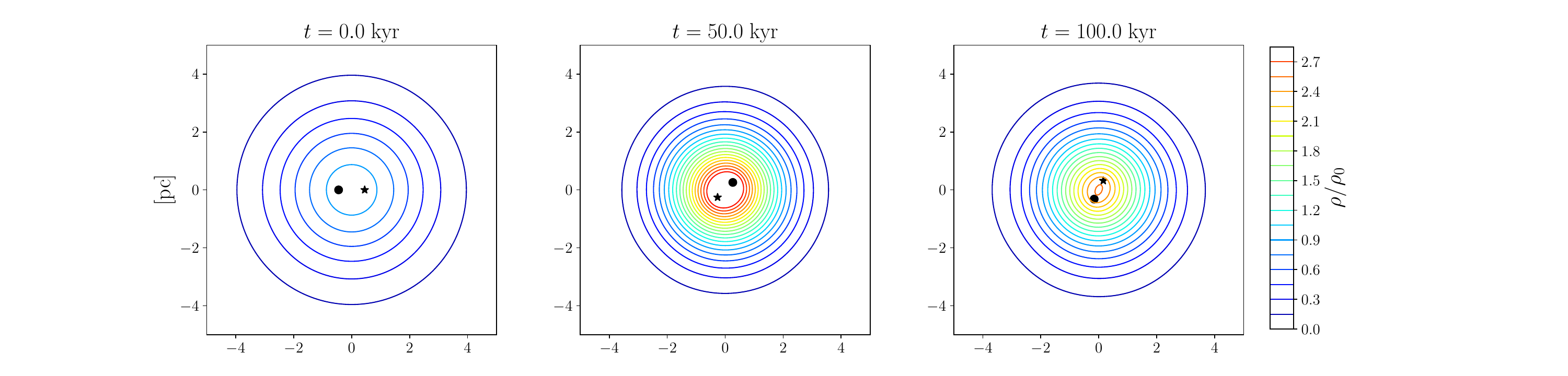}
 \caption{Time evolution of the SMBHB of mass $M_{\rm bh}=0.6\times 10^8 {\rm M}_\odot$ inside ULDM halo of mass $M_s=10^9 {\rm M}_\odot$.
 Colored lines represent the contour lines of ULDM density, where 
 $\rho_{0}$ denotes the initial central density of an isolated ground-state ULDM halo system, as
 given in  (\ref{rho0}). 
 The black dot and star in each diagram stand for our BHB embedded. See Figure~\ref{decaydata} for the orbital decay patterns. The BHs are revolving about $30$ times within several periods of central density oscillations, whose details are not fully shown in this figure.}
 \label{snapshot}
\end{figure*}

Figure~\ref{snapshot} shows three snapshots of ULDM profiles with $M_{\rm bh}=0.6\times 10^8 {\rm M}_\odot$. 
The initial profile is depicted on the left panel,  the $t=50\,{\rm kyr}$ on the middle, and the $t=100\,{\rm kyr}$ on the right. As the two BHs are revolving around each other, their movement stirs the surrounding ULDM halos and generates ULDM waves. The central region of the ULDM profile becomes elliptical, indicating a dipole-like perturbation by the SMBHB. Part of these waves escape the central region while carrying away the kinetic energy of the BHB, which is nothing but the gravitational cooling effect. 

\begin{figure*}[tb]
\centering
\includegraphics[width=0.49\textwidth]{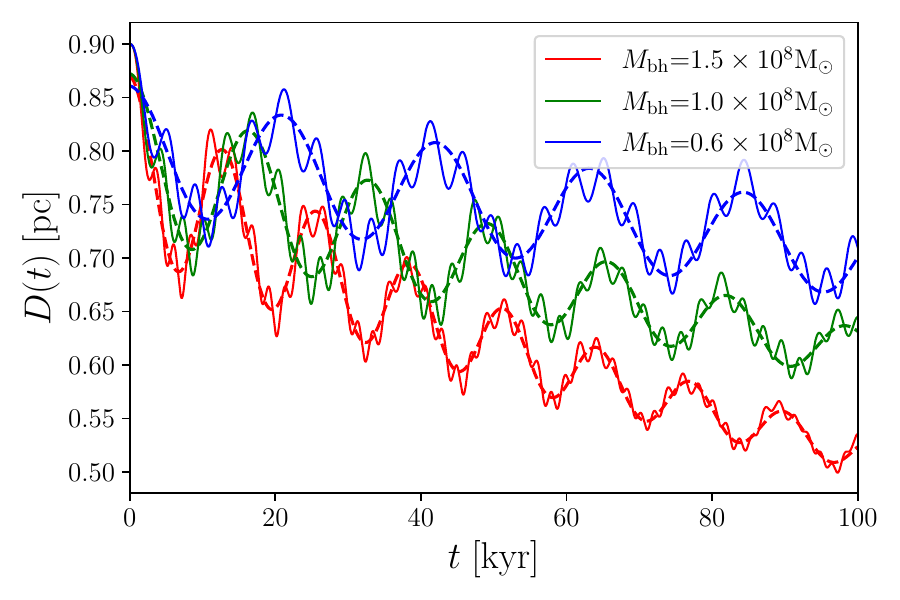}
\includegraphics[width=0.49\textwidth]{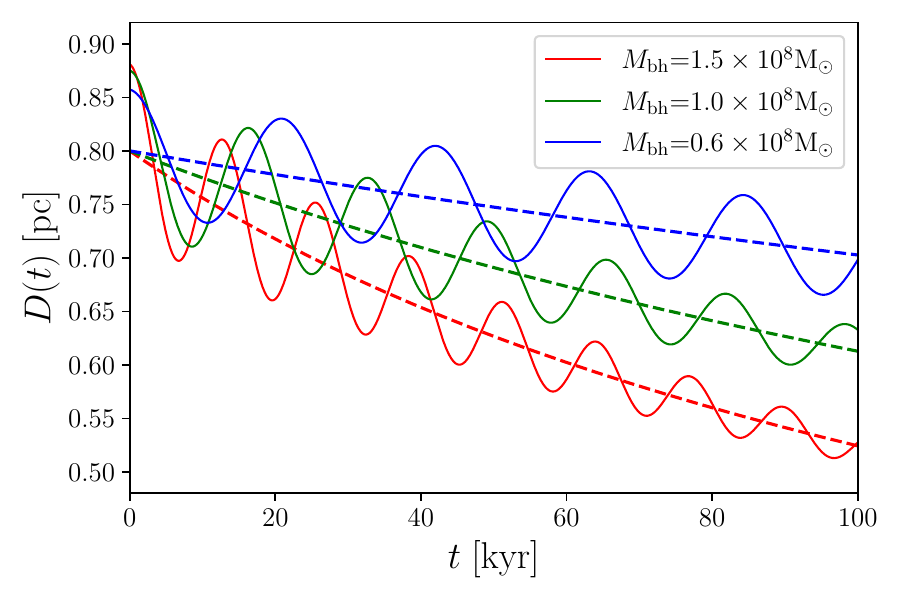}
 \caption{
  Time evolution of BHB separation $D(t)$ for $M_{\rm bh} = (0.6, 1.0, 1.5) \times 10^8 {\rm M}_\odot$ (blue, green, and red lines). Left: simulated $D(t)$ (solid) and its full fit (dotted). Right: full fit of $D(t)$ (solid) and the {\it averaged} fit $\bD(t)$ (dotted). See the texts for the definitions of full fit and averaged fit.
 }
 \label{decaydata}
\end{figure*}

The time evolution of the binary separation $D(t)$ is our main concern here and depicted on the left panel of Figure~\ref{decaydata} for $M_{\rm bh}=(0.6, \,1.0,\, 1.5)\times 10^8 \, {\rm M}_\odot$. 
Note that our initial ULDM halo configuration is not stationary since the extra interaction, due to the BHB, was initially not taken into account. The presence of BHB subsequently makes some ULDM halos attracted toward the BHs, and drives the ULDM halo system to oscillate radially together with the oscillation of the central ULDM density. This will be the major deriving force behind yet another gravitational cooling of the entire system. 

As time progresses, the initial collapse will be quickly settled down with regular oscillations. The time evolution of $D(t)$ in general exhibits the slow and fast oscillations with a slowly decaying average $\bar{D}(t)$. Firstly, the fast oscillation is simply due to the slight eccentricity of the orbit and will be ignored in the following analysis. Next, the slow oscillation mainly follows from the global quasi-normal-mode \cite{quasinormal} breathing of halo system together with the spherical central density oscillation, which is closely related to the gravitational cooling of the overall system. If there is no central BH, the time scale of this oscillation is of order of ${{\hbar }^3}/({m^3 G^2 M_s^2})$, where $M_s$ is the mass of the soliton. In our case, SMBHB can slightly change the time scale ($T_{\rm slow}$ below). On the other hand, the decaying average $\bar{D}(t)$ seems to be related to another quasi-normal mode with a time scale $\tau_g$ from the local dipole perturbation by the rotating SMBHB at the center rather than by the global oscillation of the soliton. This mode is non-spherical and can carry away the angular momentum of the SMBHB (see Sec.~\ref{sec:theoretical} for details).

\begin{figure}[tb]
\centering
\includegraphics[width=0.98\textwidth]{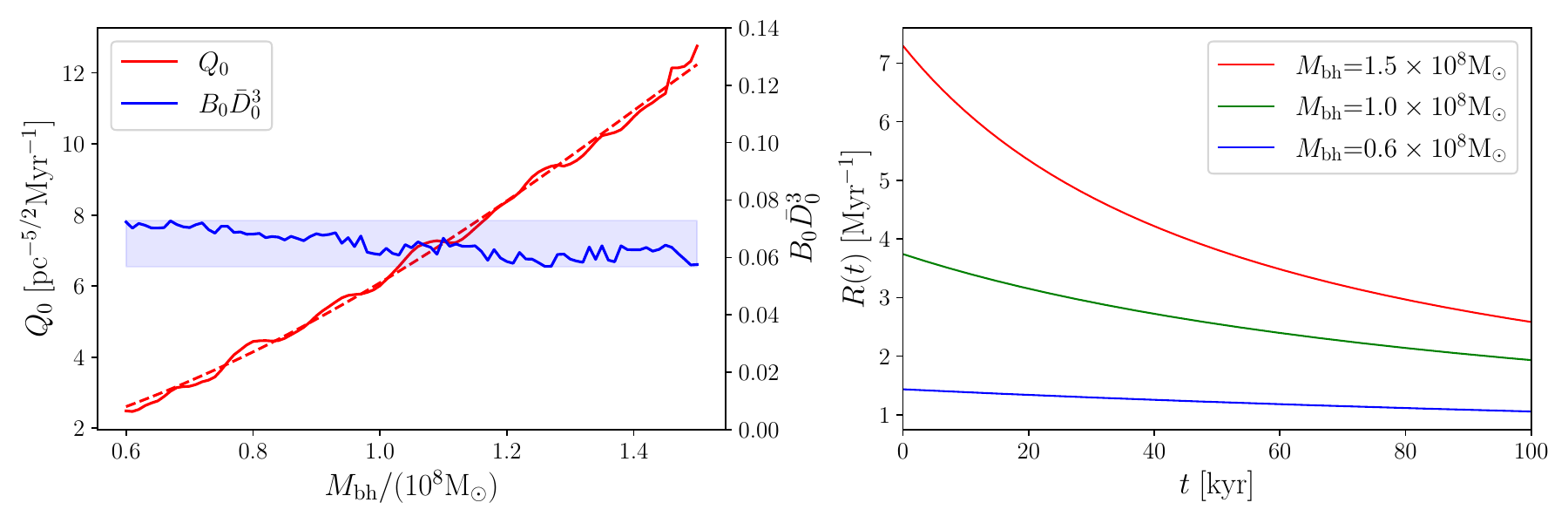}
\caption{Left: fitting parameters $Q_0$ (red line) and $B_0 \bD_0^3$ (blue line) as a function of $M_{\rm bh}$ derived from the simulations for $M_{\rm bh} = 0.6-1.5 \times 10^8 {\rm M}_\odot$. Red dotted line: the best-fit of $Q_0 (M_{\rm bh})$ from the theoretical model described in Sec.~\ref{sec:theoretical}. Light-blue shaded region emphasizes that $B_0 \bD_0^3$ is nearly constant over $M_{\rm bh}$.
Right: time evolution of the decay rate of BHB separation $R(t)$ for $M_{\rm bh} = (0.6, 1.0, 1.5) \times 10^8 {\rm M}_\odot$ (blue, green, and red lines).}
\label{decayrate}
\end{figure}

In the remainder of this section, we shall focus on extracting the {\it averaged} orbital decay function $\bar{D}(t)$, largely independent of the slow global oscillations. However, there is no well-defined procedure for such averaging, since the oscillations in Figure~\ref{decaydata} are in their nonlinear regime. Ignoring the fast oscillation, we try the following fit function 
\beq
D(t)=\bar{D}(t) +A_s \, e^{-\alpha_s t} \cos(\omega t -\phi) \, ,
\label{dfit}
\eeq
where the {\it averaged} function $\bar{D}(t)$ is  defined by 
\beq
\bD(t)=\bD_0\left(1+\frac{5}{2}Q_0\,\bD_0^{\frac{5}{2}}\,(t-t_0)\right)^{-\frac{2}{5}}
\label{dbarfit}
\eeq
with the initial time $t_0=0$ in our case (see the next section for the derivation of equation~(\ref{dbarfit})).
It is clear that $1/(Q_0 \bar{D}^{\frac{5}{2}
}_0)$ gives an initial decay time scale.

We perform the $D(t)$ fitting with the  fitting parameters,  $(Q_0,\bar{D}_0,\,A_s,\alpha_s,\, \omega,\,\phi)$. For each of $M_{\rm bh}=(0.6, \,1.0,\, 1.5)\times 10^8 \, {\rm M}_\odot$, we depict the fit function of $D(t)$ as a dotted curve on the left panel and as a solid curve on the right panel of Figure~\ref{decaydata}. We have also added $\bar{D}(t)$ as a dotted curve for the comparison on the right panel. One finds $\bar{D}_0 \simeq 0.8\, {\rm pc}$ there.
The $Q_0$ as a function of $M_{\rm bh}$ is depicted on the left panel of Figure~\ref{decayrate}.  Using the parameters and the fit function $\bar{D}(t)$, one can also obtain the decay rate given by
\beq
R(t)\equiv -\frac{{\rm d}\log \bar{D}}{{\rm d}t} = \frac{Q_0\bar{D}_0^{\frac{5}{2}}}{1+\frac{5}{2} Q_0\bar{D}_0^{\frac{5}{2}}\,(t-t_0)} \, ,
\label{drate}
\eeq
which is plotted on the right panel of Figure~\ref{decayrate} for $M_{\rm bh}=(0.6, \,1.0,\, 1.5)\times 10^8 \, {\rm M}_\odot$.

In the above fit, we also obtain the slow oscillation period $T_{\rm slow}\equiv 2\pi/\omega$, which may be further fit with the formula
\beq
    T_{\rm slow}
    = \frac{q_0 \hbar^3}{m^3 G^2 (M_s+ 2 \gamma_s M_{\rm bh})^2} \, ,
\eeq
where $q_0$ and $\gamma_s$ are the fitting parameters. We numerically find that $q_0 = 86.53 \pm 0.29$ and $\gamma_s = 1.979 \pm 0.012$. Without the BHB, this time scale is well known (see for instance \cite{Bak:2020det}). Here the main difference comes from the additional BH mass contribution weighted by the $\gamma_s$ factor. In principle, the system may involve multiple quasi-normal modes, which are characterised by the  parameters $(q_0,\, \gamma_s)$. Especially, the quasi-normal mode caused by the orbital motion of BHB may involve different parameters as discussed in the next section.

Similarly to equation~(\ref{dfit}), the relative velocity of one BH with respect to the other can be fitted with the fit function given by 
\beq
v_{\rm rel}(t)=\bar{v}_{\rm rel}(t) +\tilde{A}_s \, e^{-\tilde{\alpha}_s t} \cos(\tilde\omega t -\tilde\phi) \label{vrelfit} 
\eeq
with $\bar{v}_{\rm rel}(t)=2\sqrt{GM_{\rm bh}(1+B_0 \bar{D}^3)/(2\bar{D})}$, which is the twice of the rotation velocity of each BH including the ULDM contribution parameterized by $M_{\rm bh} B_0 \bar{D}^3=4\delta M_s$ (see the next section for the details). The fitting parameter $B_0\bD_0^3$ can be used for the consistency of our approximation in the next section and is depicted on the left panel of Figure~\ref{decayrate}.

\begin{figure}[tb]
\centering
\includegraphics[width=0.98\textwidth]{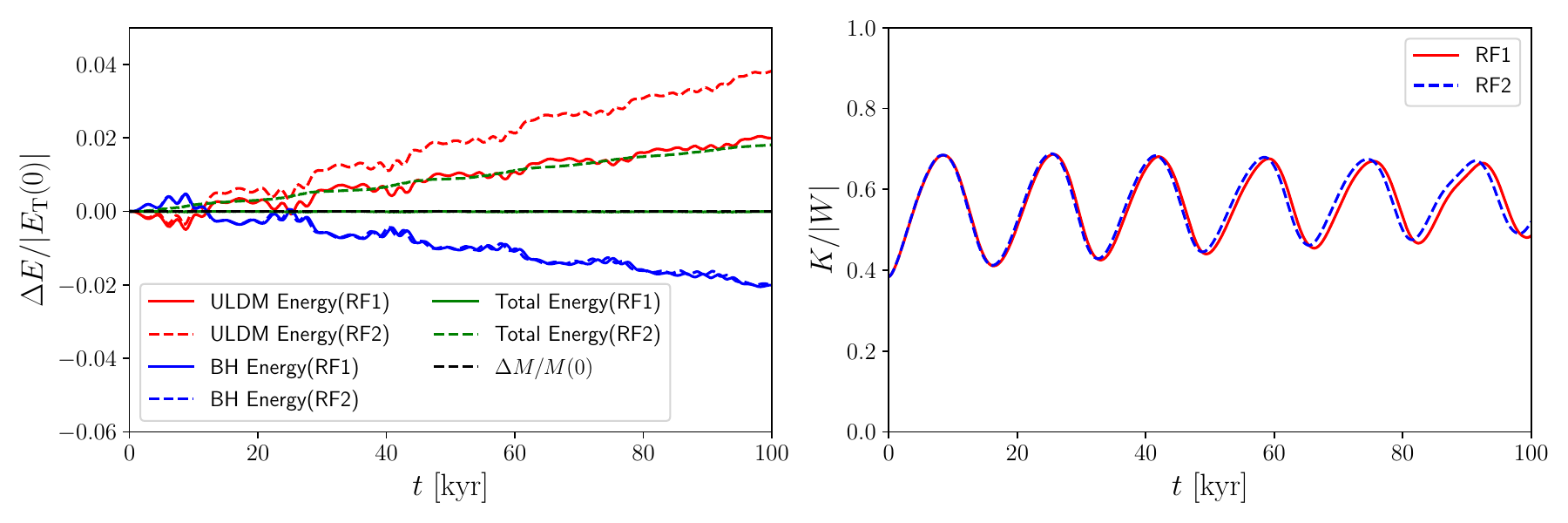}
 \caption{ Left: Time evolution of the ULDM energy, the BH energy, and the total energy for RF1 and RF2 (with $M_{\rm bh}=1.0\times 10^8 {\rm M}_\odot$). The black dashed line represents relative variation of 
 the total mass of ULDM $\Delta M/M(0)$, which is negligible.  Right: The ratio of the total kinetic energy and the absolute value of the  total potential energy.
 }
 \label{energy}
\end{figure}

To demonstrate the stability of our simulation, we present the time evolution of ULDM energy, BH energy, and total energy in the left panel of Figure~\ref{energy}. Additionally, we depict the relative variation of the ULDM mass, expressed as $\Delta M/M(0)$, where $M(0)$ denotes the initial ULDM mass. It is evident that the energy of BBHs is transferred to ULDM energy, while the total mass of ULDM remains conserved throughout the simulation. The right panel of the figure reveals that the virial ratio is about one half \cite{quasinormal}.

\section{Theoretical analysis}\label{sec:theoretical}

In this section, we provide a rough estimate of the orbital decay timescale 
of a rotating SMBHB. 
This estimate is mainly based on dimensional analysis and intended to provide  guidance for more accurate future calculations, because the system is highly nonlinear. 
Although it is difficult to clearly distinguish the contribution of dynamical friction from that of gravitational cooling, 
it turns out that, for our orbital decay problem, the dynamical friction (DF) is  dominant over the gravitational cooling (GC) effect, especially in the small separation limit. This may be argued as follows.

We begin with the case of the DF 
contribution of a BH immersed in the ULDM halo given by
\beq
F_{\rm DF}= 4\pi \rho \left(\frac{GM_{\rm bh}}{v}\right)^2 C \, ,
\label{dfriction}
\eeq 
where $\rho$ is the ULDM mass density and $C$ is the DF 
coefficient.  
For the present context, this coefficient is rather well understood. See, for instance, equation~(D8) of \cite{Hui:2016ltb} for its details. There was introduced an effective cutoff radius scale $r_{\rm eff}$ of the surrounding spherical region over which  the DF perturbation 
is effective. This scale may be taken as the smaller of the orbital radius $r=\bD/2$ and the soliton size scale. 
 In our case, since the orbital size is smaller than the  soliton size, 
we take $r_{\rm eff}=\alpha_{\rm DF} \bD/2$ where  $\alpha_{\rm DF}$ is an ${\cal O}(1)$ numerical constant which may be estimated by our simulation.
When $k\, r_{\rm eff} \ll 1$ with $k=mv/\hbar$, $C$ can be approximated by \cite{Hui:2016ltb,Wang:2022grqc}
\beq
C=\frac{1}{3} (k\,r_{\rm eff})^2 = \frac{\alpha^2_{\rm DF}}{12} \left(\frac{mv \bD}{\hbar}\right)^2
\label{C}.
\eeq
It is well known that the value of $C$ for ULDM is smaller than that for CDM. However, the wave nature (quasi-normal modes) of ULDM helps to replenish the loss cone. Figures~\ref{snapshot} and \ref{decaydata} illustrate rapid redistribution of central density of ULDM over time, unlike CDM. This might help the fast orbital decay with ULDM.

Due to the DF, the black holes lose their angular momentum
$L_{\rm bh}=M_{\rm bh}rv$.  From equations~(\ref{dfriction}), (\ref{C}), and $v\simeq \sqrt{GM_{\rm bh}/2\bar{D}}$,
we obtain the decay time scale \cite{Hui:2016ltb,Buehler:2022tmr},
\beqa
\label{taurho}
\tau &\equiv& \frac{L_{\rm bh}}{rF_{\rm DF} }
=\frac{1}{8\pi C \rho} \sqrt{\frac{M_{\rm bh}}{2G\bar{D}^3}}
=\frac{3 {\hbar}^2}{\sqrt{2} \pi  G^{3/2}} \frac{1}{\alpha_{\rm DF}^2  m^2\rho 
\sqrt{\bar{D}^5 {M_{\rm bh}}}  } \nonumber \\
&=& \left( \frac{8.6\times 10^{-3}
}{\alpha_{\rm DF}^2} \right) \left(\frac{m}{10^{-21}\rm eV}\right)^{-2} \left(\frac{M_{\rm bh}}{10^8 \mathrm{M}_\odot}\right)^{-\frac{1}{2}
} 
\left(\frac{\rho}{10^5 \mathrm{M}_\odot/\mathrm{pc}^3}\right)^{-1}
\left(\frac{\bar{D}}{\mathrm{pc}}\right)^{-\frac{5}{2}
} 
{\rm Gyr} \, ,
\eeqa
where we later obtain $\alpha_{\rm DF}\simeq 0.424$ 
from our simulation in Figure~\ref{decayrate} (see 
equation~
(\ref{evoleq}))
and $\rho$ is the average DM density
at $r=\bar{D}/2$.

\begin{figure}[tb]
\centering
\includegraphics[width=0.8\textwidth]{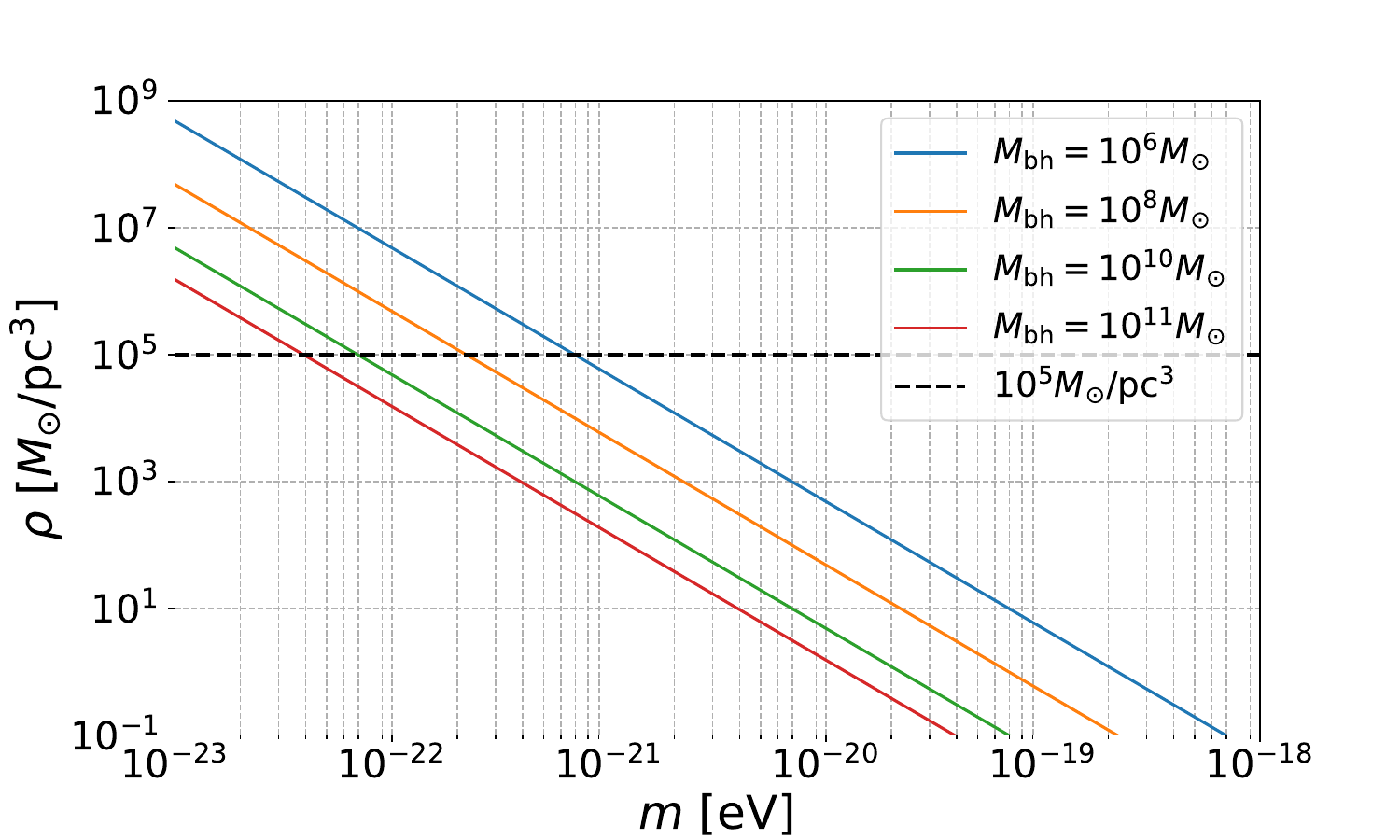}
 \caption{The thick 
oblique lines represent
 the theoretical boundary where $\tau=1\,\rm Gyr$
 for $M_{\rm bh}=(10^6,10^8,10^{10},10^{11}) \mathrm{M}_\odot$, respectively, from the top to the bottom. (See equation (\ref{taurho}).) The dashed horizontal line indicates the observational bound for the central dark matter density $\rho=10^5 \mathrm{M}_\odot/\mathrm{pc}^3$ at $r\simeq 1\,\mathrm{pc}$. The 
 separation is taken as $\bar{D}=1\,\mathrm{pc}$ and $\alpha_{\rm DF}=0.424$.
 For each $M_{\rm bh}$,
 only the parameter region below the horizontal line and over the corresponding thick line is allowed in our model.
 }
 \label{bound}
\end{figure}

The above equation shows that 
DF by ULDM can lead to an
efficient
orbital decay of BHBs within $\rm Gyr$ for moderate parameter ranges.
To show that this rapid decay is not  due to unnatural high dark matter density, we plot the allowed region in the parameter plane
$(m,\rho)$ for $\tau<1  \,\rm Gyr$
in Figure~\ref{bound} with $\alpha_{\rm DF}=0.424$.
From the graph one can see that the heavier
$m$ is, the lower $\rho$ can be, and
there are plausible parameter regions with $m\gtrsim 10^{-22}\rm eV$ with
relatively low DM density compatible with observations.

For the lowest BH mass observed ($M_{\rm bh}\simeq 10^6 \mathrm{M}_\odot$),
which can lead to the longest
$\tau$,
one can get $m > 6.9 \times 10^{-22}\rm eV$, if $\rho < 10^5 \mathrm{M}_\odot/\mathrm{pc}^3$. 
Conversely, 
if we choose $m < 6.9\times 10^{-22}\rm eV$ and $\rho < 10^5 \mathrm{M}_\odot/\mathrm{pc}^3$, 
we can find a bound for the black hole mass
$M_{\rm bh} > 10^6 \mathrm{M}_\odot$.
This means that BHBs with mass smaller than $10^6 \mathrm{M}_\odot$
cannot merge efficiently to grow to supermassive black holes with  $m\simeq 10^{-22}\rm eV$, which is often suggested to solve the small scale issues. This fact might explain why supermassive black holes typically have mass  $M_{\rm bh} > 10^6 \mathrm{M}_\odot$.

There is another constraint on $m$  if we use the soliton
density  $\tilde{M}_s/2/(4 \pi r_h^3/3)$ for $\rho$. Here, $\tilde{M}_{ s}=M_{ s}+2\gamma M_{\rm bh}$ is an effective core mass determining the frequency of the quasi-normal mode   from
the ULDM halo mass combined with the BHB mass weighted by a factor $\gamma$ and  
$r_{{h}}$ is the half-mass radius  
of the ULDM halo  given by
$f_0{{\hbar }^2}/({m^2 G{\tilde{M}}_s})$ with $f_0=3.925$ (see for instance \cite{Hui:2016ltb}).
Then the above equation can be rewritten as
\beqa
\tau &=& \frac{4 \sqrt{2} {f_0}^3 {\hbar}^8}{G^{9/2}} \frac{1}{\alpha_{\rm DF}^2 m^8 \tilde{M}_s^4 \sqrt{{M_{\rm bh}}\bar{D}^5}} \nonumber \\
&=& \left(\frac{2.72\times 10^{-4}
}{\alpha_{\rm DF}^2}\right)  \left(\frac{m}{10^{-21}\rm eV}\right)^{-8} \left(\frac{M_{\rm bh}}{10^8 \mathrm{M}_\odot}\right)^{-\frac{1}{2}
} 
\left(\frac{\tilde{M}_s}{10^9 \mathrm{M}_\odot}\right)^{-4}
\left(\frac{\bar{D}}{\rm pc}\right)^{-\frac{5}{2}
}
{\rm Gyr} \, , 
\label{tau2}
\eeqa
which is sensitive to $m$.
 The condition $\tau<1\,\rm Gyr$ 
now gives a bound $m > 9.1\times 10^{-23}\rm eV$ if we
use the observational upper bounds $M_{\rm bh} \lesssim 10^{11} \mathrm{M}_\odot$
and 
$\tilde{M}_s \lesssim 10^{10} \mathrm{M}_\odot$~\cite{2016MNRAS.456L.109K}.
Remarkably, although our analysis 
about DF is not directly related to the small-scale problems, the bounds for $m$ above are quite similar to the typical values 
that resolve 
the small-scale issues \cite{fuzzy}.

DF contribution may be compared to the one from the 
GC effect which is estimated below.
As was mentioned already, 
the ULDM waves generated by the BHB carry away the momentum and orbital energy of the BHs. 
The corresponding GC 
timescale of the quasi-normal mode 
with SMBHB is given by 
$\tau_g ={{\hbar }^3}/({m^3 G^2\tilde{M}^2_s})$  \cite{quasinormal}.
Since the perturbation has a local origin, this factor 
$\gamma$ is in general slightly different from the global one ($\gamma_s$) in $T_{\rm slow}$ and 
will be determined separately below. 

Each 
BH 
is carrying 
some extra amount of ULDM distribution  that is gravitationally bound to the BH.
We focus on the excess amount above the orbital average for a given orbital radius and call
this mass $\delta M$ bound to the BH. As we are considering a BHB with a separation $\bD$, this density perturbation 
should be dipole-like in the small separation limit, and may be estimated as
${\delta M}=\alpha 
M_{\rm bh}({\pi \bD}/{r_{{h}}})$
at the linear order,
where $\alpha(M_{\rm bh}/M_s)$
is a 
dimensionless numerical coefficient. 

We suggest that the magnitude of the momentum change 
of the ULDM halo should be balanced with 
that of $\delta M$, and then with the corresponding BH since $\delta M$ is bound to the BH.
We  expect that mainly ULDM waves carry the increased momentum 
of the halo. 
For each revolution, if there were no dissipation, the BH would return to the original starting position and then the effective dissipative force, due to the GC 
effect, will be weighted by the cooling faction ${\Delta }{\tau }_{{c}}/\tau_g$ where ${\Delta }{\tau }_{{c}} \equiv {\pi \bD}/{v}$
is approximately the orbital period of the BHs.
This fraction can be also interpreted in the following way: We are looking to calculate the effective average force over the orbital period of the BHBs. Given that the time scale for gravitational cooling is about $\tau_g$, we anticipate that a fraction, represented as ${\Delta }{\tau }_{{c}}/\tau_g$, of the soliton mass ${\tilde{M}}_s$ could be removed as ULDM waves due to the cooling throughout this period.
Then, the effective force between the ULDM waves outgoing and  $\delta M$ is roughly
$\bar{F}_{\rm GC}\propto G 
({\tilde{M}}_s {\Delta }{\tau }_{{c}}/\tau_g )\delta M/{r^2_{{h}}}$.
Note that we are estimating an average force for a specific moment in time, not solely for a single period. To determine the total momentum exchange, one must integrate the average force over the full duration of time. 
Thus, the effective force due to the GC 
effect
is  roughly
\beq
F_{\rm GC}
=K_0 
\left( \frac{{\Delta }{\tau }_{{c}}}{\tau_g } \right) 
\left( \frac{G
{\tilde{M}}_s \delta M}{r^2_{{h}}} \right) \, ,
\label{deltap}
\eeq
 where $K_0$ is a numerical constant, 
 and 
 ${{\Delta }{\tau }_{{c}}}/{\tau_g }$ 
 represents the above cooling fraction 
 of the averaged force $\bar{F}_{\rm GC}$. 

The ratio of $F_{\rm GC}$ over $F_{\rm DF}$ at given time may be straightforwardly evaluated  as
\beq
\frac{F_{\rm GC}}{F_{\rm DF}}=
\left( \frac{8\pi^2 \alpha K_0}{\alpha^2_{\rm DF}} \right) \sqrt{\frac{2f_0\bD\, \tilde{M}_s^3}{r_h\,M_{\rm bh}^3}} \,,
\label{ratio}
\eeq
where we again approximate the density of ULDM halo as $\rho=(\tilde{M}_s/2)/(4\pi r_h^3/3)$. Assuming the orbital decay in our simulation is solely from the 
DF effect, we find below that $\alpha_{\rm DF}\sim 0.424
$ that is indeed of order one. 
This  implies that,  for our simulation,  the GC 
contribution can be initially at most comparable to that of the DF. 
Also, in the small separation limit, the DF 
becomes dominant over the GC 
effect.   
 Thus, from now on, we shall ignore the GC 
effect in our theoretical calculation and assume the orbital decay is mainly due to the DF. 
However, GC plays an important role during the orbital decay, as the spherical quasi-normal modes with the time scale $T_{\rm slow}$ can rapidly refill the phase space of ULDM that was removed by the SMBHB. ($T_{\rm slow}$ is much shorter than the decay time scale $\tau$.) 
On the contrary, collisionless CDM experiences loss cone depletion due to its long relaxation time as mentioned earlier.
It is one reason why ULDM better explains the rapid orbital decay, despite having a smaller $C$ compared to CDM. 
Another problem of CDM is that the heat from DF of CDM can destroy CDM spikes \cite{2003MNRAS.339..949R}.

 Now, we compare our approximate calculation of $\tau$ with the numerical results.
From equation~(\ref{dfriction}), the infinitesimal orbital velocity change due to the DF  is given by ${\rm d}v= {\rm d}t \, (F_{\rm DF}/M_{\rm bh})$.
Assuming a circular orbit, the orbital velocity of 
BHs depends on both the BH 
mass ($M_{\rm bh}$) and the  ULDM mass ($\delta M_s$) enclosed within the orbit, with  Kepler's law:
$v = \sqrt{G\tilde{M}_{\rm bh}/2\bD}\,$ where $\tilde{M}_{\rm bh} \equiv M_{\rm bh} + 4\, \delta M_s$.
The enclosed mass within $\bD /2$ is
$\delta M_s 
= 4\pi\rho \,(\bD/2)^3/3 =\eta\, 
m^6_{21} \tilde{M}^4_s\bD^3 
=(B_0 \bD^3) M_{\rm bh} /4$, where $\eta$ is a constant and 
$B_0$ is introduced in the previous section.
From now on, we assume $B_0 \bD^3 \ll 1$ so that its 
contribution in $\tilde{M}_{\rm bh}$ is ignored in the  computation below.
We also checked that our  
simulations
in Sec. \ref{sec:numerical}
indeed satisfy this condition.
Then, $v\propto \bD^{-\frac{1}{2}
}$ and 
$\left|{{\rm d}
\bD}/{\bD}\right|=2\left|{{\rm d} 
v}/{v}\right|
$.

Finally, one obtains  an approximate differential equation for $\bD(t)$,
\beq
-\frac{{\rm d} \log \bD}{
{\rm d}t}=\frac{\alpha^2_{\rm DF}  m^8 G^{\frac{9}{2}} {\tilde{M}_s}^4M^{\frac{1}{2}}_{\rm bh}  
}{2\sqrt{2}{f_0}^3 {\hbar}^8 
} \bD^{\frac{5}{2}} =Q_0 \bD^{\frac{5}{2}} ,
\label{evoleq}
\eeq
where $Q_0\equiv \kappa\, m^8_{21}\tilde{M}_s^4 M^{\frac{1}{2}}_{\rm bh}$ 
with a parameter $\kappa\equiv {\alpha^2_{\rm DF}  m^8  G^{\frac{9}{2}} }/(2\sqrt{2}{f_0}^3 {\hbar}^8 m^8_{21})$.
 This equation  is perfectly consistent with the two scaling relations of our Schr\"odinger-Poisson system and 
 may be integrated 
 leading to the solution in equation~ 
(\ref{dbarfit}).

Based on the above analysis, we fit $Q_0$ in Figure~\ref{decayrate} by assuming the quadratic form $\kappa=\kappa_0 \left[ 1+10\, \delta_1 (M_{\rm bh}/M_s)
+100\,\delta_2 (M_{\rm bh}/M_s)^2 \right]$ 
with fitting parameters $(\gamma, \kappa_0,\delta_1,\delta_2)$. This fit of $Q_0$ is depicted on the left panel of Figure~\ref{decayrate} by a dotted curve. One finds
$\gamma = 2.192 \pm 0.034$, 
$\kappa_0 = (1.324 \pm 0.029) \times 10^{-43}\,{\rm kyr^{-1} pc^{-\frac{5}{2}} \mathrm{M}_\odot^{-\frac{9}{2}}}$
 with negligibly small $\delta_1$ and $\delta_2$. 
Then, $\kappa\simeq \kappa_0$
and it corresponds to $\alpha_{\rm DF}\simeq 0.424.$

We also obtain
\beq
Q_0
 =1.324\left(\frac{\tilde{M}_s}{10^9  \mathrm{M}_\odot}\right)^4 \left(\frac{M_{\rm bh}}{10^8 \mathrm{M}_\odot}\right)^{1/2}\left(\frac{m}{10^{-21} {\rm eV}}\right)^8 
 {\rm Myr^{-1} \, pc^{-\frac{5}{2}}} \, ,
 \label{Q0}
\eeq
where the number 
$1.324$
is from the fitting.
The initial decay rate, given by $ R(t_0) = \left| {\rm d} \log \bar{D} / {\rm d}t 
\right| \simeq Q_0 \bar{D}_0^{\frac{5}{2}} $ in equation~(\ref{drate}), can be related to the decay scale $\tau$ in equation~(\ref{tau2}) as $ \tau \simeq 2/R(t_0) 
= {2}/({Q_0 \bar{D}_0^{\frac{5}{2}}})$.
For example, Figure~\ref{decayrate} shows that $2/R(t_0) \simeq 0.617~\text{Myr}$
for $M_{\rm bh}=10^8 \,\mathrm{M}_\odot$ from our simulation, where 
$\tilde{M}_{s}=M_{ s}+2\gamma M_{\rm bh}\simeq 1.438 \times 10^9 \,\mathrm{M}_\odot$.
This time scale from simulation is comparable to the  theoretical prediction $\tau = 0.47 ~\text{Myr}$ in equation~(\ref{tau2}) with simple assumptions $\alpha_{\rm DF} = 1$, $\tilde{M}_{s}\simeq M_s = 10^9\, \mathrm{M}_\odot$ and $\bar{D}_0 = 0.8\,\mathrm{pc}$, thereby supporting our analysis.
To check a full decay time, we need to do a simulation up to $\bar{D}(t)\sim 0.01\,\mathrm{pc}$, which is beyond the scope of this work.

\section{Summary}\label{sec:summary}

In this letter, we demonstrate through numerical simulations that ULDM waves generated by rotating BHB in ULDM halos can cause rapid orbital decay
and give a hint to the final parsec problem.
This phenomenon is rather unexpected because the orbital decay of BHB in ULDM halos is generally considered to be slow due to the weak dissipation effects~\cite{PhysRevD.105.083008}. 
The orbital decay arises mainly from 
DF and GC of ULDM halos interacting with the BHB where DF becomes dominant over GC effect in the small separation limit of the BHB.
{One advantage of ULDM is that it is free from the loss cone problem, unlike collisionless CDM or  stars. This is because ULDM has a wavelike nature, which helps it rapidly refill the phase space due to GC.
Figure~\ref{snapshot} shows that the central density of ULDM may even increase as the separation of the black holes decreases unlike CDM. Investigating the influence of the ULDM waves on nearby stars might offer another potential solution to the loss-cone problem \cite{Bromley:2023yfi}.}

These findings have implications for future gravitational wave probes like LISA or NANOGrav, as  ULDM waves can change the orbital 
properties of the 
SMBHBs and influence gravitational waves generated by them, especially when their separation is small. To further validate our results, higher-resolution simulations are planned for future studies.

\begin{acknowledgments}
We would like to thank Sangnam Park for the support of our numerical simulation. 
This research was supported by Basic Science Research Program through the National Research Foundation of Korea (NRF) funded by the Ministry of Education (2018\-R1\-A6\-A1\-A06024977) for HK, DB and IP.
DB was also supported in part by NRF Grant RS-2023-00208011.
SEH was supported by the projects ``Understanding Dark Universe Using Large Scale Structure of the Universe''  and ``Research on the Principle of Accelerated Expansion of the Universe'' funded by the Ministry of Science.
This work was supported by the Korea Institute of Science and Technology Information, through the KREONET network and the supercomputing resources. This work was also supported by the
UBAI computing resources at the University of Seoul.
\end{acknowledgments}

\end{document}